\pdfoutput=1
\documentclass[lettersize, journal]{IEEEtran}
\usepackage{amsmath,amsfonts}
\usepackage{algorithmic}
\usepackage{algorithm}
\usepackage{array}
\usepackage[caption=false,font=normalsize,labelfont=sf,textfont=sf]{subfig}
\usepackage{textcomp}
\usepackage{stfloats}
\usepackage{url}
\usepackage{verbatim}
\usepackage{graphicx}
\usepackage{cite}
\usepackage{amssymb,bm}
\usepackage{setspace,stfloats}
\usepackage[colorlinks,linkcolor=blue]{hyperref}
\usepackage{ulem}
\usepackage{color}
\usepackage{bigstrut, multirow, multicol, rotating}
\definecolor{mygray}{RGB}{160, 160, 160}

\usepackage[switch]{lineno}

\begin{document}


\title{ESTformer: Transformer Utilizing Spatiotemporal Dependencies for Electroencephalogram Super-resolution}

\author{Dongdong~Li,
        Zhongliang Zeng,
        Zhe~Wang$^{*}$,
        Hai~Yang 
\IEEEcompsocitemizethanks{\IEEEcompsocthanksitem DongDong Li, Zhongliang Zeng, Zhe Wang, and Hai Yang are with the East China University of Science and Technology.}

\thanks{$^*$Corresponding author. E-mail: wangzhe@ecust.edu.cn(Z. Wang).}}

\maketitle

\begin{abstract}
  Towards practical applications of Electroencephalography (EEG), lightweight acquisition devices garner significant attention.
  However, EEG channel selection methods are commonly data-sensitive and cannot establish a unified sound paradigm for EEG acquisition devices.
  Through reverse conceptualisation, we formulated EEG applications in an EEG super-resolution (SR) manner, but suffered from high computation costs, extra interpolation bias, and few insights into spatiotemporal dependency modelling.
  To this end, we propose ESTformer, an EEG SR framework that utilises spatiotemporal dependencies based on the transformer.
  ESTformer applies positional encoding methods and a multihead self-attention mechanism to the space and time dimensions, which can learn spatial structural correlations and temporal functional variations.
ESTformer, with the fixed mask strategy, adopts a mask token to upsample low-resolution (LR) EEG data in the case of disturbance from mathematical interpolation methods.
  On this basis, we designed various transformer blocks to construct a spatial interpolation module (SIM) and a temporal reconstruction module (TRM).
  Finally, ESTformer cascades the SIM and TRM to capture and model the spatiotemporal dependencies for EEG SR with fidelity.
  Extensive experimental results on two EEG datasets show the effectiveness of ESTformer against previous state-of-the-art methods, demonstrating the versatility of the Transformer for EEG SR tasks.
  The superiority of the SR data was verified in an EEG-based person identification and emotion recognition task, achieving a 2\% to 38\% improvement compared with the LR data at different sampling scales.
\end{abstract}


\section{Introduction\label{sec:1}}
\IEEEPARstart{W}{ith} the rapid advancements in artificial intelligence and neuroscience, electroencephalography (EEG)-based brain--computer interface technology has enabled a wide range of applications, including biometrics \cite{liPersonalizedFederatedContinual2023}, healthcare \cite{wangPerformanceEnhancementP3002021, KHARE2023110858}, and intelligent control systems \cite{wangLinkingAttentionBasedMultiscale2021, LI2023110179}, offering users novel and transformative life experiences.
For practical applications, EEG acquisition devices must meet key requirements such as user comfort and affordability \cite{sainiDSCNNCAUDeepLearningBasedMental2023}. Consequently, design paradigms for lightweight and portable EEG devices have attracted considerable attention. Researchers have explored various design strategies to meet these demands, such as enhancing and reducing the number of electrodes. A common approach is to employ EEG channel selection techniques to eliminate redundant electrodes \cite{alyasseriPersonIdentificationUsing2020}.
However, these methods are highly data-sensitive and often yield varying channel combinations depending on the manual features extracted from the data acquired using different protocols and devices \cite{ashenaeiStableEEGBasedBiometric2022}. In addition to channel selection, EEG applications face a similar challenge: the need for diverse methods tailored to different protocols and devices \cite{wangApproachOnevsRestFilter2020, wangDiverseFeatureBlend2020, xuMotorImageryDecoding2023}. This complexity arises from intrinsic difficulties in processing data from heterogeneous devices and protocols \cite{liuHybridConvolutionalNeural2024}.
Thus, although a unified paradigm for EEG devices is crucial for advancing EEG applications, it remains difficult to achieve this using current channel-selection approaches alone.

In general, diverse techniques can be applied to process data acquired through lightweight EEG devices to enhance the performance of subsequent tasks, such as data augmentation to expand dataset samples \cite{baoDataAugmentationEEGBased2021}, recover poor segments in some electrodes based on effective electrodes \cite{saba-sadiyaEEGChannelInterpolation2020}, and extract better EEG representations by self-supervised learning with a random mask strategy \cite{kostasBENDRUsingTransformers2021}.
Nevertheless, these studies mainly focus on the question of how to better represent EEG in the given data dimensions and can hardly reconcile into a unified electrode system design.
In addition, the implicit neural information in the uncovered electrode areas remains to be explored for advanced applications, especially for low-spatial-resolution EEG data.

In computer vision, deep-learning methods \cite{hanTimelyDetectionSkin2024, zhangDeepLearningOutline2024} are commonly applied to preprocessed data using mathematical techniques such as artefact removal \cite{razmjooyImperialistCompetitiveAlgorithmBased2017}, image segmentation \cite{razmjooyHybridNeuralNetwork2018}, and noise reduction \cite{xuComputeraidedDiagnosisSkin2020},  \cite{caiBreastCancerDiagnosis2021}. 
Consequently, researchers have emphasised these upstream preprocessing techniques to assess their impact on downstream performance.
A notable area within these techniques is image reconstruction, which investigates the inherent characteristics of data in low-level tasks, such as super-resolution (SR) \cite{sunTwoStageDeepSingleImage2023}, to address ill-posed problems and establish a more robust foundation for higher-level tasks \cite{yooRethinkingDataAugmentation2020}. 
Advancements in SR technology have provided valuable insights into recovering EEG data from unobserved channels.
In general, there are two methodological approaches for integrating the SR reconstruction technology into EEG.
One approach involves an end-to-end strategy that employs deep-learning methods directly to upsample low-spatial-resolution EEG data \cite{corleyDeepEEGSuperresolution2018}.
An alternative approach adopts a stage-to-stage methodology that combines mathematical interpolation techniques with deep-learning methods for data refinement \cite{hanFeasibilityStudyEEG2018}.

Despite their demonstrated feasibility, existing EEG SR methods face three significant limitations that hinder their effectiveness:

1) High computational cost: Both end-to-end and stage-to-stage approaches frequently rely on convolutional neural network (CNN)-based generative adversarial networks (GANs) or other complex frameworks, which results in excessive computational overhead and limits real-time applications \cite{corleyDeepEEGSuperresolution2018}.

2) Extra-interpolation bias: In stage-to-stage frameworks, traditional mathematical interpolation techniques, often used as preprocessing steps, can introduce substantial bias rather than capture the true underlying characteristics of complex EEG data. This bias can negatively affect the generalisation performance of subsequent deep-learning models, particularly when handling challenging or noisy EEG signals \cite{kangDistortionsEEGInterregional2015}.

3) Lack of spatiotemporal dependencies: From a spatiotemporal perspective, end-to-end methods typically do not account for the physical distances between electrodes during interpolation, which limits their ability to capture spatial relationships. In addition, effectively modelling the temporal dependencies inherent in EEG data remains an open challenge that has not been adequately addressed in current models \cite{svantessonVirtualEEGelectrodesConvolutional2021, tangDeepEEGSuperresolution2023}.

To overcome these challenges, our emphasis is on end-to-end frameworks to obviate the need for mathematical interpolation methods.
The backbone of the framework should proficiently model data in both spatial and temporal dimensions.
Therefore, we propose ESTformer, which is a framework based on a transformer \cite{vaswaniAttentionAllYou2017} with a fixed-mask strategy.

In the development of end-to-end frameworks for EEG SR, an intuitive approach is to employ a fixed-mask strategy, in which the model is trained to reconstruct the data in masked channels based on the information available from other channels. 
However, only a few studies have explored fixed-mask strategies for EEG applications. 
Instead, a random mask strategy is commonly adopted, particularly in self-supervised pretraining paradigms \cite{heMaskedAutoencodersAre2022}.
Although the random mask strategy offers marginally better performance by presenting a model with diverse views of the same data, it may conflict with the specific requirements of the EEG spatial SR task. 
By generating random masks across epochs, the random mask strategy risks exposing the model to nearly all channels, which could lead to unintended leakage of unseen channel representations, particularly in the EEG SR task described in \cite{tangDeepEEGSuperresolution2023}.
Based on this, we hypothesised that the fixed-mask strategy holds significant potential for EEG SR reconstruction, guiding our framework design.

In recent studies, the mask strategy has commonly been accompanied by an encoder--decoder framework based on a transformer, which is compact and easy to train.
The Transformer is mainly implemented using a multihead self-attention (MSA) mechanism with positional encoding, which can explore spatial structural correlations between different channels of EEG data, conforming to the SR task.
The application of the transformer in the field of long-term sequence data also demonstrated its potential for analysing and modelling temporal variations in EEG data with high temporal resolution \cite{zhouInformerEfficientTransformer2021}.
The Transformer, which is primarily composed of MSA and linear layers, can establish long-term dependency relationships and support parallel computing, which can alleviate the time-consuming issues associated with previous methods.

In summary, the transformer-based framework has the potential to effectively model the functional connectivity of different brain regions at a low time cost.
Therefore, we propose a spatial interpolation module (SIM) and temporal reconstruction module (TRM) based on a transformer to capture spatial structural correlations and temporal functional variations in EEG, respectively.
The proposed ESTformer, composed of SIM and TRM, can utilise spatial and temporal dependencies to efficiently achieve high-quality signal spatial dimension expansion, thereby improving the performance of downstream tasks in EEG, such as person identification and emotion recognition.

The main contributions of this study are as follows.

1) This study proposes ESTformer, a transformer-based EEG SR framework with less computation.
We apply the fixed-mask strategy for the EEG SR task to upsample the EEG data without introducing extra bias from mathematical interpolation methods.
ESTformer, establishing the mapping between low-resolution (LR) and SR data, can be adapted to diverse lightweight acquisition devices.

2) To better model data in space and time dimensions, we apply the MSA to introduce space-wise MSA (SSA) and time-wise MSA (TSA).
Leveraging the SSA and TSA, we introduced the SSA block (SSAB) and TSA block (TSAB).
Considering the spatiotemporal dependency, we propose the cross-attention block (CAB) and further build SIM and TRM with 3D spatial and 1D temporal positional encoding methods.

3) Extensive experiments were conducted to verify the EEG SR performance of the proposed ESTformer.
The downstream tasks of person identification and emotion recognition were studied to discuss the effectiveness of EEG SR data in terms of common handcrafted features in different frequency bands, compared to the LR data and ground truth (GT) data.

The remainder of this paper is organised as follows: Sec.\ref{sec:2} discusses related work, Sec.\ref{sec:3} presents the proposed ESTformer framework, Sec.\ref{sec:4} demonstrates the experimental results including the EEG SR reconstruction task and the EEG downstream tasks, and Sec.\ref{sec:5} concludes the paper.

\section{Related Work \label{sec:2}}
\subsection{EEG Interpolation and SR \label{sec:2-1}}
The EEG source is subject to strong volume conduction distortion owing to the conduction characteristics of the scalp and the “many-to-many” relationship between scalp electrodes and brain sources.
To model the human brain, a volume conduction model can be established using electromagnetics with a spatial mathematical foundation \cite{hassanElectroencephalographySourceConnectivity2018}.
Thus, traditional interpolation methods for EEG are based on the strength of spatial mathematical interpolation analysis.
Nouira et al. \cite{nouiraEEGPotentialMapping2014} proposed barycentric and spline interpolation methods for EEG.
Khouaja et al. \cite{khouajaEnhancingEEGSurface2016} combined the Kalman filter with spherical spline interpolation (SI) \cite{freedenSphericalSplineInterpolation1984}.
These mathematical interpolation methods are highly dependent on the position of the electrodes and are sensitive to the calculation of distances between electrodes \cite{courellisEEGChannelInterpolation2016}. However, only the type of electrode system can be accessed instead of the precise position of the electrodes in many EEG datasets and practical applications, introducing some deviations caused by different human head shapes and wearable modes.

Consequently, researchers have begun to use deep-learning methods in an end-to-end manner for EEG interpolation and SR by establishing encoder--decoder architectures and GAN primarily based on CNN.
Corley et al. \cite{corleyDeepEEGSuperresolution2018} applied a Wasserstein GAN (WGAN) to an EEG SR with better stability.
Saba-Sadiya et al. \cite{saba-sadiyaEEGChannelInterpolation2020} proposed a deep encoder--decoder network to interpolate artefacts in EEG in general and transfer learning settings.
Svantesson et al. \cite{svantessonVirtualEEGelectrodesConvolutional2021} proposed three encoder--decoder networks to reconstruct EEG data under different SR settings.
Sun et al. \cite{sunDesignVirtualBCI2023} simply applied the Informer \cite{zhouInformerEfficientTransformer2021} to various EEG channel interpolation tasks.

Recent studies have mainly focused on a stage-to-stage manner, tending to use mathematical interpolation methods for interpolation, and then using other deep-learning models to improve the interpolation results.
Han et al. \cite{hanFeasibilityStudyEEG2018} interpolated unseen channels by averaging their neighbouring channels and proposed a deep CNN-based model for data augmented by adding noise.
Panwar et al. \cite{panwarGeneratingEEGSignals2019} proposed a conditional WGAN with a gradient penalty to interpolate single-channel EEG data based on bicubic interpolation and bilinear weight initialisation.
Tang et al. \cite{tangDeepEEGSuperresolution2023} studied brain structure and functional connectivity and adopted suitable graph neural networks and CNNs to obtain EEG SR data based on preliminary interpolated EEG data using the SI method \cite{freedenSphericalSplineInterpolation1984}, and then compared the SR EEG data with the original LR EEG data in case studies.
Although these stage-to-stage methods that combine mathematical interpolation methods with deep-learning models have achieved better performance, they rely on the modelling ability of mathematical interpolation methods, thereby introducing significant bias in challenging data and weakening the generalisation ability of subsequent deep-learning models \cite{ouahidiSpatialGraphSignal2023}.

In recent years, there have been studies on brain magnetic resonance image (MRI) SR based on transformers, demonstrating the modelling and generation capabilities of brain image data \cite{zhang3dCrossScaleFeature2022}.
Although the feasibility of the self-attention mechanism has been verified in EEG \cite{liSpatialfrequencyConvolutionalSelfattention2022}, it mainly serves EEG classification-related tasks \cite{luoDualBranchSpatioTemporalSpectralTransformer2023}.
When using a transformer, there are still few insights into electrode spatial position encoding specifically for EEG \cite{arjunIntroducingAttentionMechanism2021} and how MSA represents the implicit information of brain region connectivity \cite{wangTransformersEEGBasedEmotion2022}.

In summary, there is still room for exploration to address the ill-posed problem of transformers in EEG-related generative tasks, as little research can be found in the literature \cite{sunDesignVirtualBCI2023}.
Therefore, we attempted to explore the transformer capability of EEG SR to alleviate the limitations of the existing methods explored in this context.

\subsection{Mask Strategy for EEG \label{sec:2-3}}
In 2019, Devlin et al. \cite{devlinBERTPretrainingDeep2019} proposed BERT, a bidirectional masked language model based on a transformer \cite{vaswaniAttentionAllYou2017}, which demonstrated remarkable performance in language understanding and suggested a random mask strategy for self-supervised learning to explore the intrinsic representations of data.
Sequentially, wav2vec 2.0 \cite{baevskiWav2vecFrameworkSelfSupervised2020} was proposed for audio signal learning tasks.
He et al. \cite{heMaskedAutoencodersAre2022} proposed MAEs based on a vision transformer (ViT) \cite{dosovitskiyImageWorth16x162020} and discussed three mask strategies: block-wise sampling, grid-wise sampling, and random sampling.
Among these, the random sampling mask strategy achieved remarkable image reconstruction quality, even at a high masking rate of 75\%.
MAEs, which serve as a strong motivation for researchers, have been applied to other domains such as graph data \cite{houGraphMAE2DecodingEnhancedMasked2023}.

These studies provide evidence for the feasibility of leveraging the transformer and random mask strategy for self-supervised learning in different data modalities; thus, existing studies on EEG pay more attention to the random mask strategy.
Inspired by wav2vec 2.0 \cite{baevskiWav2vecFrameworkSelfSupervised2020}, Kostas et al. \cite{kostasBENDRUsingTransformers2021} proposed the pretraining model BENDR for EEG.
Similarly, recent methods reconstructed EEG data for sleep-detection tasks \cite{chienMAEEGMaskedAutoencoder2022} and reconstructed handcrafted features extracted from EEG data for emotion-recognition tasks \cite{liMultiviewSpectralSpatialTemporalMasked2022}.

In summary, transformer-based frameworks with a random mask strategy have shown promising experimental results for exploring the latent representations of EEG.
However, despite the image reconstruction ability of the grid-wise sampling mask strategy in MAEs \cite{heMaskedAutoencodersAre2022}, few studies have discussed a transformer-based framework with a fixed-mask strategy for EEG tasks.
Therefore, we adopted a fixed-mask strategy conforming to the requirements of the EEG SR task, which may provide a way to overcome the challenges in EEG SR and support lightweight devices for EEG applications.

\begin{figure*}[!htbp]
  \centering
  \includegraphics[width=1\textwidth]{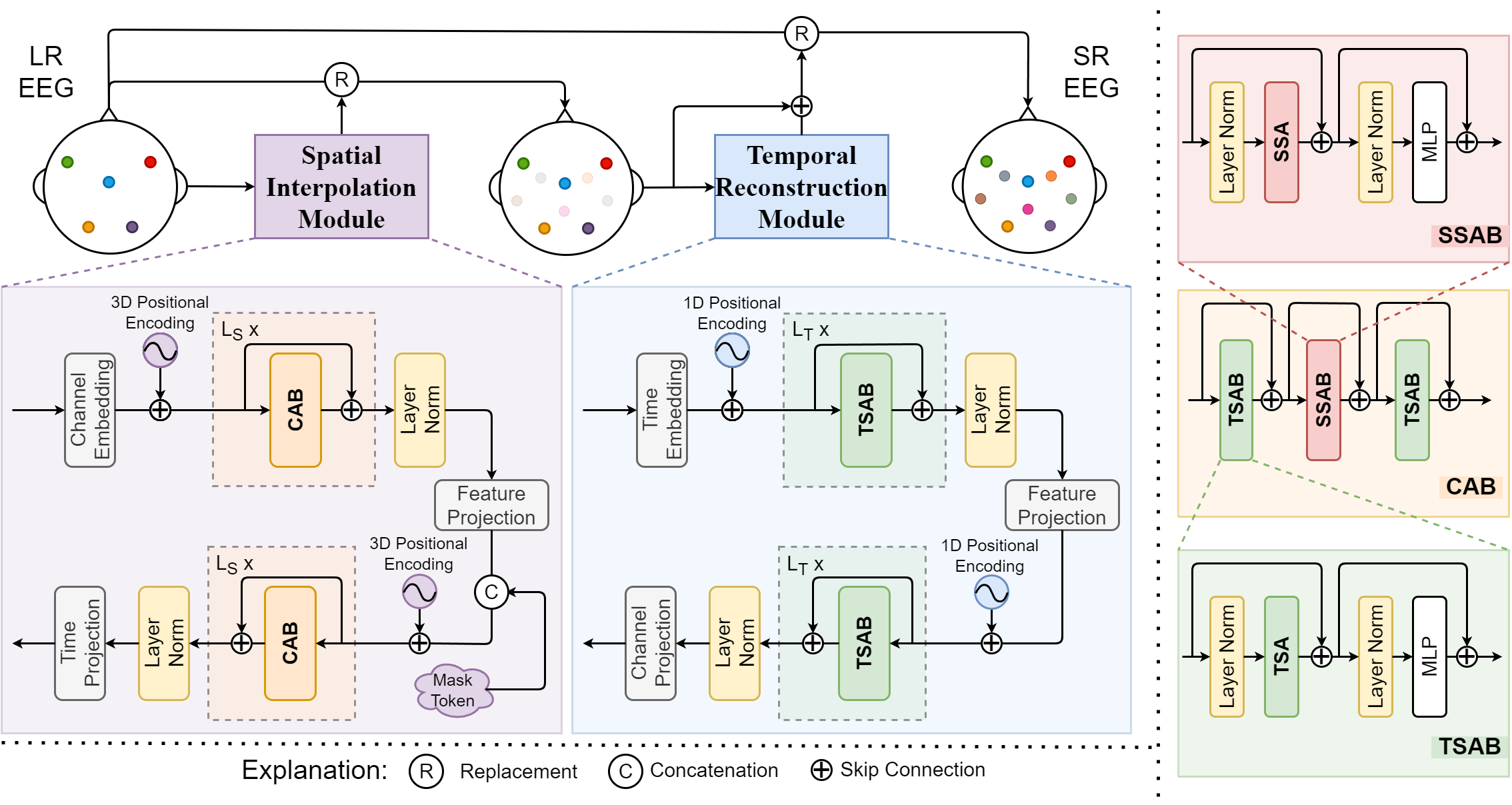}
  \caption{Overview of the ESTformer framework.
    ESTformer is composed of SIM and TRM.
    LR EEG data first pass through SIM to reconstruct missing channels with 3D spatial positional encodings.
    In SIM, CAB is composed of SSAB and TSAB, modelling spatiotemporal dependencies, to better obtain SR EEG data.
    Then, SR EEG data pass through TRM, as an auxiliary module, to refine temporal relationships by TSAB with 1D temporal positional encodings.
    In SSAB and TSAB, SSA and TSA are depicted in \textbf{Fig.\ref{fig:SSA}} and \textbf{Fig.\ref{fig:TSA}}.
    Linear layers are leveraged for Embedding and Projection layers.}
  \label{fig:ESTformer}
\end{figure*}

\begin{figure}[!htpb]
  \centering
  \includegraphics[width=1\columnwidth]{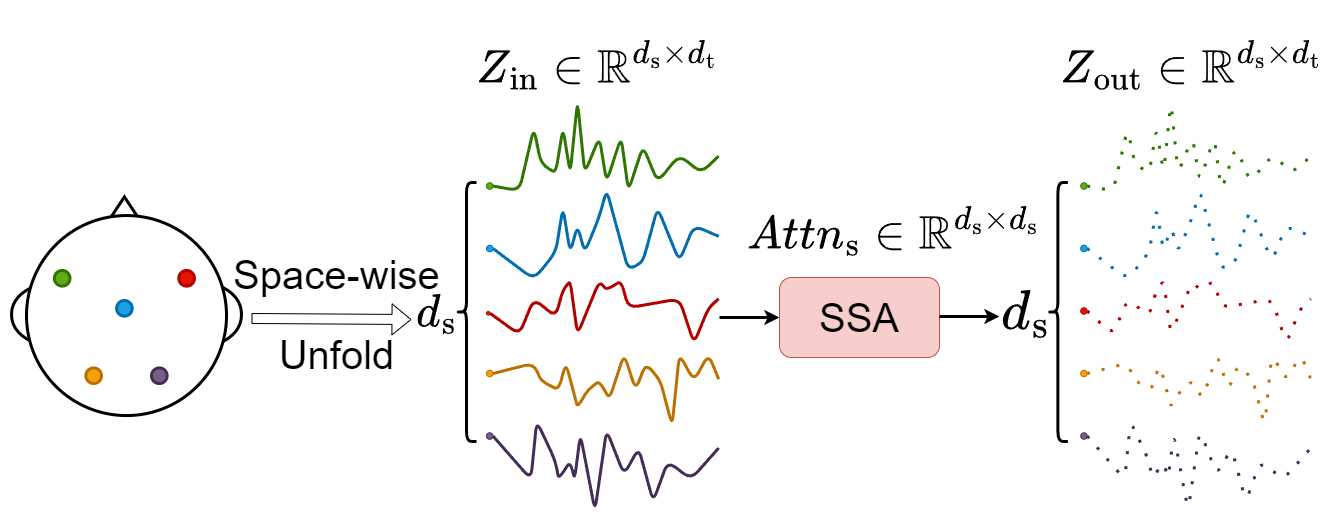}
  \caption{SSA pipeline in SSAB, modelling spatial relationships. Unfolding the input according to electrode channels and regard time-wise data as feature dimensions.}
  \label{fig:SSA}
\end{figure}

\begin{figure}[!htpb]
  \centering
  \includegraphics[width=1\columnwidth]{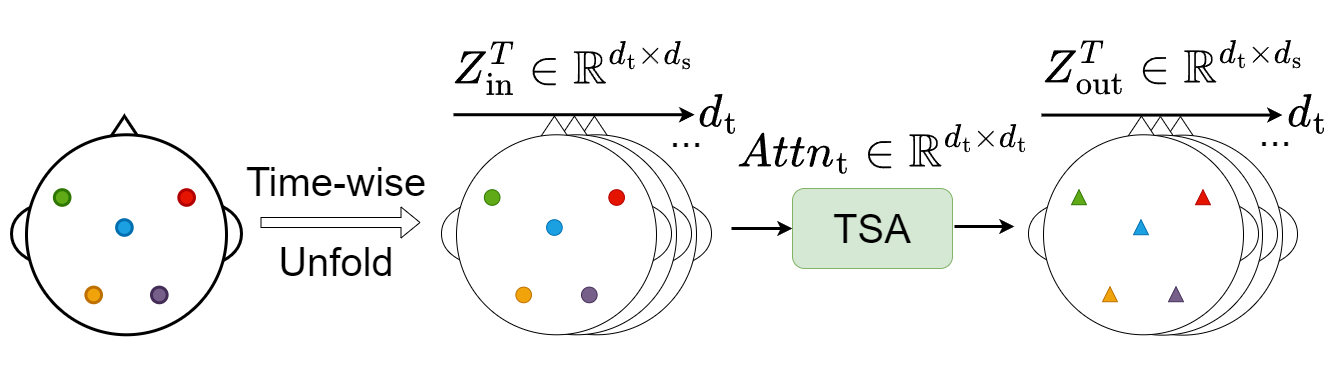}
  \caption{TSA pipeline in TSAB, modelling temporal relationships. Unfolding the input according to time samples and regard space-wise data as feature dimensions.}
  \label{fig:TSA}
\end{figure}

\section{Methodology \label{sec:3}}

In the EEG SR task (specific to spatial SR), given LR EEG data denoted as $X_{\rm{LR}}\in \mathbb{R}^{C_{\rm{LR}}\times T}$, the objective is to reconstruct EEG SR data denoted as $X_{\rm{SR}}\in \mathbb{R}^{C_{\rm{SR}}\times T}$ by interpolating the missing channels of EEG data denoted as $X_{\rm{Mask}}\in \mathbb{R}^{C_{\rm{Mask}}\times T}$.
$C_{\rm{SR}}$ and $C_{\rm{LR}}$ denote the numbers of electrode channels in $X_{\rm{SR}}$ and $X_{\rm{LR}}$ respectively.
$C_{\rm{Mask}}$ denotes the number of missing channels that must be interpolated (i.e. channels masked in the original data).
$T$ denotes the number of time-sampling points for $X_{\rm{SR}}$ and $X_{\rm{LR}}$, indicating the same temporal resolution.

In this study, we propose ESTformer, an EEG SR framework utilising spatial and temporal dependencies based on a transformer, to alleviate the high time cost and inadequate capacity of modelling challenging EEG data in existing methods.
The proposed ESTformer is shown in \textbf{Fig.\ref{fig:ESTformer}}, composing the SIM and TRM.
In SIM, we adopt CAB based on two types of transformer blocks.
One is the space-wise self-attention block (SSAB), which captures spatial dependency using space-wise self-attention (SSA).
The other is the time-wise self-attention block (TSAB) to capture the temporal dependency by time-wise self-attention (TSA), which is also adopted to construct the TRM.

In summary, given the hypothesis that exploring temporal variations and spatial correlations in EEG data can help utilise the inherent spatiotemporal dependencies for EEG SR, we applied MSA to both the spatial and temporal dimensions, introducing space-wise MSA (SSA) and time-wise MSA (TSA) for EEG data in \ref{sec:3-1}.
We then construct SSAB and TSAB, which are combined in the CAB in \ref{sec:3-2}.
Based on CAB, we built the SIM in \ref{sec:3-3}.
Regarding the high temporal resolution characteristics of the EEG data, we built the TRM based on the TSAB in \ref{sec:3-4}.
Finally, in \ref{sec:3-5}, because stacking blocks mainly focus on high-frequency features \cite{liuImprovingPixelbasedMIM2023}, we establish an ESTformer framework with dense residual connections to learn multilevel features based on SIM and TRM.

\subsection{Space- and Time-wise Self-Attention \label{sec:3-1}}
Inspired by Crossformer \cite{zhangCrossformerTransformerUtilizing2022}, MSA \cite{vaswaniAttentionAllYou2017} was applied to EEG data in both spatial and temporal dimensions.

As shown in \textbf{Fig.\ref{fig:SSA}}, given the spatiotemporal latent variable $Z_{\rm{in}}$, the output variable $Z_{\rm{s}}$, representing spatial structural correlations, can be obtained by Eq. \eqref{eq:SSA}.
Similarly, as shown in \textbf{Fig.\ref{fig:TSA}}, given the $Z_{\rm{in}}$, the output variable $Z_{\rm{t}}$, representing temporal functional variations, can be obtained by Eq. \eqref{eq:TSA}.
In Eq. \eqref{eq:SSA} and Eq. \eqref{eq:TSA}, $Z_{\rm{in}}$, $Z_{\rm{s}}$, $Z_{\rm{t}} \in \mathbb{R}^{d_{\rm{s}}\times d_{\rm{t}}}$, where $d_{\rm{s}}$ and $d_{\rm{t}}$ denote the sizes of the spatial and temporal dimensions of the features respectively.

\begin{equation}
  Z_{\rm{s}}=\mathbf{MSA}\left(Z_{\rm{in}}\right)
  \label{eq:SSA}
\end{equation}

\begin{equation}
  Z_{\rm{t}}^T=\mathbf{MSA}\left(Z_{\rm{in}}^T\right)
  \label{eq:TSA}
\end{equation}

By estimating the computational complexity using floating point operations (FLOPs), the FLOPs calculation formulas for the SSA and TSA can be expressed as Eq. \eqref{eq:F_SSA} and Eq. \eqref{eq:F_TSA} respectively.

\begin{equation}
  \mathit{FLOPs}_{\rm{SSA}}=4{d_{\rm{s}}}{d_{\rm{t}}}^2+2{d_{\rm{s}}}^2{d_{\rm{t}}}
  \label{eq:F_SSA}
\end{equation}

\begin{equation}
  \mathit{FLOPs}_{\rm{TSA}}=4{d_{\rm{t}}}{d_{\rm{s}}}^2+2{d_{\rm{t}}}^2{d_{\rm{s}}}
  \label{eq:F_TSA}
\end{equation}

Existing deep-learning-based EEG interpolation methods primarily use 2D convolution layers, and the FLOPs calculation formula for this approach can be expressed as \eqref{eq:F_Conv}.

\begin{equation}
  \mathit{FLOPs}_{\rm{Conv2d}}=C_{\rm{in}}C_{\rm{out}}{k_{\rm{s}}}{k_{\rm{t}}}{d_{\rm{s}}}{d_{\rm{t}}}
  \label{eq:F_Conv}
\end{equation}

In Eq. \eqref{eq:F_Conv}, $C_{\rm{in}}$ and $C_{\rm{out}}$ denote the number of convolution channels for the input data and the output data respectively. $k_{\rm{s}}$ and $k_{\rm{t}}$ denote the spatial and temporal sizes of the convolution kernels, respectively.

According to EEGSR-GAN \cite{corleyDeepEEGSuperresolution2018}, assuming that $d_{\rm{s}}=64$, $d_{\rm{t}}=1600$, $k_{\rm{s}}=33$, $k_{\rm{t}}=1$, $C_{\rm{in}}=128$, $C_{\rm{out}}=128$, the calculations can be obtained that $\mathit{FLOPs}_{\rm{SSA}}\approx0.67\rm{G}$, $\mathit{FLOPs}_{\rm{TSA}}\approx0.35\rm{G}$, $\mathit{FLOPs}_{\rm{Conv2d}}\approx55.36\rm{G}$.
Hence, EEG interpolation models constructed based on the MSA can significantly reduce the high time cost in existing EEG interpolation models based on two-dimensional convolutional neural networks (2D-CNNs) \cite{corleyDeepEEGSuperresolution2018, hanFeasibilityStudyEEG2018}.

In the implementation, considering the 3D spatial positional encoding of the EEG electrodes, the SSA sets the number of heads for the MSA to three.
However, considering the 1D temporal positional encoding of EEG sampling points, the TSA sets the number of heads for the MSA to one (not using the multihead mechanism).

\subsection{Transformer Block Design \label{sec:3-2}}
The key to the EEG spatial SR task lies in modelling the spatial dependencies of EEG and capturing spatial structural correlations.
Inspired by the transformer \cite{vaswaniAttentionAllYou2017} and ViT \cite{dosovitskiyImageWorth16x162020}, the proposed SSA is combined with a multilayer perceptron (MLP) to form the SSAB, as shown in Eq. \eqref{eq:SSAB_SSA} - \eqref{eq:SSAB_MLP}, where $\mathbf{LN}$ represents Layer Normalisation, denoted as Layer Norm in \textbf{Fig.\ref{fig:ESTformer}}.

\begin{equation}
  Z_{\rm{temp}}=\mathbf{SSA}\left.\left(\mathbf{LN}\left.\left(Z_{\rm{in}}\right.\right)\right.\right)\ +\ Z_{\rm{in}}
  \label{eq:SSAB_SSA}
\end{equation}

\begin{equation}
  Z_{\rm{S}}=\mathbf{MLP}\left.\left(\mathbf{LN}\left.\left(Z_{\rm{temp}}\right.\right)\right.\right)\ +\ Z_{\rm{temp}}
  \label{eq:SSAB_MLP}
\end{equation}

The high temporal resolution characteristics of EEG data require models that utilise temporal dependencies to capture temporal functional variations and improve the EEG SR. performance.
Similarly, the TSAB can be formed by combining the MLP with the TSA, as shown in Eq. \eqref{eq:TSAB_TSA} - \eqref{eq:TSAB_MLP}.

\begin{equation}
  Z_{\rm{temp}}^T=\mathbf{TSA}\left.\left(\mathbf{LN}\left.\left(Z_{\rm{in}}^T\right.\right)\right.\right)\ +\ Z_{\rm{in}}^T
  \label{eq:TSAB_TSA}
\end{equation}

\begin{equation}
  Z_{\rm{T}}^T=\mathbf{MLP}\left.\left(\mathbf{LN}\left.\left(Z_{\rm{temp}}^T\right.\right)\right.\right)\ +\ Z_{\rm{temp}}^T
  \label{eq:TSAB_MLP}
\end{equation}

Because the MLP is based on linear layers, its parameter size and computational complexity are positively correlated with the feature dimensions.
Specifically, the SSAB regards the time dimension in EEG data as a feature, whereas the TSAB regards the spatial dimension in EEG data as a feature.
Consequently, for EEG data with low spatial resolution but high temporal resolution, the TSAB owns a smaller parameter size and requires a lower computational cost than SSAB.

Inspired by CAT \cite{linCATCrossAttention2022}, we leverage the TSAB as an auxiliary module for the SSAB.
On the one hand, we suppose that combining TSAB to explore temporal variations can strengthen the representation ability of the SSAB to capture better spatial structural correlations.
Thus, the CAB, composed of the SSAB and TSAB through residual connections, is proposed to construct the SIM.
After integrating the temporal-dimension features, the spatial dependencies of the EEG are modelled to preliminarily reconstruct $X_{\rm{SR}}$, denoted as ${\hat{X}}_{\rm{Mask}}\in \mathbb{R}^{C_{\rm{Mask}}\times T}$.
On the other hand, we directly constructed the TRM based on the TSAB by connecting $X_{\rm{LR}}$ and ${\hat{X}}_{\rm{Mask}}$ according to the electrode channels to obtain $X_{\rm{temp}}\in \mathbb{R}^{C_{\rm{SR}}\times T}$.
The temporal dependencies of $X_{\rm{temp}}$ are modelled to rearrange ${\hat{X}}_{\rm{Mask}}$ in $X_{\rm{temp}}$, thereby enhancing the reconstruction effect of the EEG spatial SR and obtaining the final $X_{\rm{SR}}$.

The implementations of SSAB, TSAB, and CAB are depicted in the right part of \textbf{Fig.\ref{fig:ESTformer}}.
The details of the SIM and TRM can be found in Sec.\ref{sec:3-3} and Sec.\ref{sec:3-4}.

\subsection{Spatial Interpolation Module\label{sec:3-3}}
The MAEs \cite{heMaskedAutoencodersAre2022} demonstrated powerful image reconstruction capabilities.
We propose a SIM with a similar architecture.

The approximate 3D spatial positions of electrodes can be obtained according to the international 10--20 system.
Based on the classical Sincos absolute positional encoding method \cite{biAccurateMediumrangeGlobal2023}, we adopt a 3D approach to encode the position information on the x-, y-, and z-axes separately, and integrate them to obtain the final 3D spatial positional encoding, as shown in Eq. \eqref{eq:PEj_sin} - \eqref{eq:PE3D}.

\begin{equation}
  \mathit{PE}_{\left(\mathit{pos}_{\left(j\right)},2i\right)}=\sin\left(\frac{\mathit{pos}_{\left(j\right)}}{{10000}^{{2i}/{d_{\rm{model}}}}}\right)
  \label{eq:PEj_sin}
\end{equation}

\begin{equation}
  \mathit{PE}_{\left(\mathit{pos}_{\left(j\right)},2i+1\right)}=\cos\left(\frac{\mathit{pos}_{\left(j\right)}}{{10000}^{{2i}/{d_{\rm{model}}}}}\right)
  \label{eq:PEj_cos}
\end{equation}

\begin{equation}
  \mathit{PE}_{\left(j\right)}=\mathbf{Concat}\left(\mathit{PE}_{\left(\mathit{pos}_{\left(j\right)},2i\right)},\ \mathit{PE}_{\left(\mathit{pos}_{\left(j\right)},2i+1\right)}\right)
  \label{eq:PEj}
\end{equation}

\begin{equation}
  \mathit{PE}_{\rm{3D}}=\mathbf{Concat}\left(\mathit{PE}_{\left(x\right)},\ \mathit{PE}_{\left(y\right)},\ \mathit{PE}_{\left(z\right)}\right)
  \label{eq:PE3D}
\end{equation}

Where ${pos}_{\left(j\right)}$ denotes the coordinate position on the j-axis in 3D space, $i$ denotes the dimension, $d_\mathit{model}$ denotes the dimension of the positional encoding ${PE} _{\left(j\right)}$ on the j-axis, and the concatenated 3D spatial position encoding is obtained as $\mathit{PE}_{\rm{3D}}$.

The CAB-based SIM can alternately learn and fuse the temporal and spatial dependencies of EEG data.
Because multilevel feature concatenation helps model convergence and alleviates the problem of focusing too much on detailed representations, residual connections are applied to multilevel CAB.
A fixed-mask strategy is used to satisfy the requirements of the EEG SR task.

The main process of SIM is illustrated on the left side of \textbf{Fig.\ref{fig:ESTformer}}.
First, the given $X_{\rm{LR}}$ is embedded in the electrode channel representations through linear layers combined with $\mathit{PE}_{\rm{3D}}$.
Leveraging the proposed CAB with residual connections, multilevel feature representation $Z_{\rm{SIM}}\in \mathbb{R}^{C_{\rm{LR}}\times d_{\rm{t}}}$ can be learned from $X_{\rm{LR}}$.
$Z_{\rm{SIM}}$ is then concatenated with the initialised mask token $Z_{\rm{Mask}}\in \mathbb{R}^{C_{\rm{Mask}}\times d_{\rm{t}}}$ as the upsampling manner, and combined with $\mathit{PE}_{\rm{3D}}$ again.
Using the same number of layers $L_{\rm{S}}$, the multilevel CAB, with the assistance of $\mathit{PE}_{\rm{3D}}$, enables $Z_{\rm{Mask}}$ to learn the spatial structural correlations of $Z_{\rm{SIM}}$ in a high-dimensional feature space, resulting in the preliminary reconstruction ${\hat{X}}_{\rm{Mask}}$.

\subsection{Temporal Reconstruction Module\label{sec:3-4}}
We assume that SIM disrupts the temporal information of the original data during the reconstruction process because of the embedding layer, resulting in interpolated data that primarily focus on the data content rather than maintaining an aligned temporal relationship with the real data.
Therefore, we propose the TRM.

Because modelling the temporal information is necessary, the spatial dimension of the electrode channels is regarded as the feature dimension.
Similar to BERT\cite{devlinBERTPretrainingDeep2019}, the classic sine--cos absolute positional encoding method is used to encode the 1D temporal position of the time series ${PE} _{\rm{1D}}$ (i.e. $\mathit{PE}_{\left(t\right)}$, where t denotes the time axis).

The TSAB-based TRM serves as an auxiliary module for the final temporal adjustment, avoiding excessive parameter growth.
Similar to SIM, residual connections are applied to handle multilevel EEG features from the TSAB.
Unlike SIM, because the input and output data are of the same dimension in TRM, no additional mask strategy, not to mention the mask token, is required.

The main process of TRM is depicted in the middle of \textbf{Fig.\ref{fig:ESTformer}}.
Firstly, based on the connection of $X_{\rm{LR}}$ and ${\hat{X}}_{\rm{Mask}}$ according to the order of electrode channels, the obtained $X_{\rm{temp}}\in \mathbb{R}^{C_{\rm{SR}}\times T}$ is embedded for temporal variations through linear layers and combined with $\mathit{PE}_{\rm{1D}}$.
Leveraging the proposed TSAB with residual connections, the multilevel feature representation $Z_{\rm{TRM}}\in \mathbb{R}^{d_{\rm{s}}\times T}$ can be learned from $X_{\rm{temp}}$.
$Z_{\rm{TRM}}$ is then combined with $\mathit{PE}_{\rm{1D}}$ again.
Using the same number of layers $L_{\rm{T}}$, the multilevel TSAB, with the assistance of $\mathit{PE}_{\rm{1D}}$, enables ${\hat{X}}_{\rm{Mask}}$ to learn the temporal variations of $X_{\rm{LR}}$ in the high-dimensional feature space $Z_{\rm{TRM}}$, reconstructing ${\hat{X}}_{\rm{Mask}}$ in the time domain.

\subsection{ESTformer Framework\label{sec:3-5}}
The overall ESTformer framework is illustrated in \textbf{Fig.\ref{fig:ESTformer}}.
Given the low-spatial-resolution EEG data $X_{\rm{LR}}$, SIM is used to capture the spatial dependencies and model the structural correlations of the EEG, resulting in a preliminary reconstruction ${\hat{X}}_{\rm{Mask}}$ for the unseen/masked electrode channels.
To learn fully from the original $X_{\rm{LR}}$, we concatenate ${\hat{X}}_{\rm{Mask}}$ and $X_{\rm{LR}}$ based on the positions of the electrode channels.
The TRM then learns the temporal dependencies and reconstructs the ${\hat{X}}_{\rm{Mask}}$ concerning temporal variations.
Similarly, the corresponding part of the output from the TRM is replaced with the original $X_{\rm{LR}}$ to obtain the final EEG SR data $X_{\rm{SR}}$.

Reconstruction losses are commonly used for EEG SR tasks, such as the mean squared error (MSE) and mean absolute error (MAE) calculated in the time domain.
In addition, EEG research often employs frequency-domain analysis \cite{liMultiviewSpectralSpatialTemporalMasked2022}.
However, measuring the distance between the reconstructed and real data in the frequency domain helps alleviate the issue of missing high-frequency information in image reconstruction tasks \cite{jiangFocalFrequencyLoss2021}.
In this study, the discrete Fourier Transform (DFT) is used to calculate the MSE (FMSE) between $X_{\rm{SR}}$ and the real EEG data$ X_{\rm{GT}}\in \mathbb{R}^{C_{\rm{SR}}\times T}$ in the frequency domain, as shown in Eq. \eqref{eq:FMSE}.

\begin{equation}
  \mathcal{L}_{\rm{FMSE}}=\sum_{m=1}^{M}\left|\mathcal{F}\left(X_{\rm{GT}}^m\right)-\mathcal{F}\left(X_{\rm{SR}}^m\right)\right|^2
  \label{eq:FMSE}
\end{equation}

where $M$ denotes the number of unseen electrode channels, $X_{\rm{GT}}^m$ and $X_{\rm{SR}}^m$ respectively denote the data of the m-th unseen channel in $X_{\rm{GT}}$ and $X_{\rm{SR}}$, and $\mathcal{F}\left(\cdot\right)$ represents the complex number form of the data by the DFT.

We combine the loss functions in both the time and frequency domains in an auto-weighted manner \cite{cipollaMultitaskLearningUsing2018} thus, the weights between different loss functions are learned by the deep-learning model.
The final loss function is given in Eq. \eqref{eq:Loss}.

\begin{align}
  \begin{split}
    \mathcal{L} &=\frac{1}{2\sigma_1^2}\mathcal{L}_{\rm{FMSE}}+\frac{1}{2\sigma_2^2}\mathcal{L}_{\rm{MAE}}+\log{\sigma_1\sigma_2} \\
    &=\frac{1}{2\sigma_1^2}\sum_{m=1}^{M}\left|\mathcal{F}\left(X_{\rm{GT}}^m\right)-\mathcal{F}\left(X_{\rm{SR}}^m\right)\right|^2 \\
    & ~ ~ ~ ~ +\frac{1}{2\sigma_2^2}\sum_{m=1}^{M}\left|X_{\rm{GT}}^m-X_{\rm{SR}}^m\right| +\log{\sigma_1\sigma_2}
  \end{split}
  \label{eq:Loss}
\end{align}

where $\sigma_1$ and $\sigma_2$ are learnable weight-related parameters for FMSE and MAE, respectively, and are updated according to the model-training process.
The selected optimiser iteratively minimises $\mathcal{L}$ until the gradients converge to an optimal or a suboptimal point.

\section{Experiments \label{sec:4}}

We conducted experiments on the EEG SR task to verify the effectiveness of ESTformer on two datasets, where we compared hyperparameter settings, and analysed the effectiveness of the main modules in ESTformer.
Because the proposed ESTformer is a general approach for downstream tasks, we conducted person identification experiments on the EEG motor movement/imagery (MI/MM) dataset \cite{schalkBCI2000GeneralpurposeBraincomputer2004} and emotion-recognition experiments on the SEED dataset \cite{zhengInvestigatingCriticalFrequency2015} to further explore the superiority of EEG SR data.
The experiments were performed on a PC equipped with a CPU (Intel(R) Core(TM) i7-10700K CPU @ 3.80 GHz), RAM(32 GB), and GPU(NVIDIA GeForce RTX 3080).

\subsection{Datasets and Preprocessing}

The EEG Motor Movement/Imagery Dataset (MI/MM) \cite{schalkBCI2000GeneralpurposeBraincomputer2004} records EEG signals from 64 channels while the subjects perform different motor tasks of movement and imagery.
The electrode positions were obtained using the international 10--20 system.
The signal-sampling frequency was set to 160 Hz.
The dataset includes over 1500 EEG records, each lasting 1–2 min, obtained from 109 subjects.
Each subject participated in 14 experiments: two baseline movements (one minute with eyes open and one minute with eyes closed) and four types of real or imagined hand or foot movements (three two-minute trials for each type).

The SEED dataset \cite{zhengInvestigatingCriticalFrequency2015} selected 15 movie clips as emotional stimuli, each lasting approximately four minutes, to evoke the corresponding emotions continuously and prominently.
This dataset includes EEG and eye-tracking data from 12 subjects and EEG data from an additional three subjects.
The movie clips were used to evoke positive, neutral, and negative emotions.
After watching each clip, participants were immediately asked to complete a survey reporting their emotional responses.
During data collection, a 62-channel international 10--20 system was used for electrode placement, with a sampling frequency of 1000 Hz.
In the preprocessed version, the EEG data were downsampled to 200 Hz and filtered using a 0--75 Hz bandpass filter.

In this study, the EEG SR experimental setup followed the Deep-EEGSR \cite{tangDeepEEGSuperresolution2023}.
The preprocessed EEG signals were divided into continuous non-overlapping 10-second samples for the MI dataset and continuous non-overlapping 4-second samples for the SEED dataset.
For each dataset, 80\% of the samples were used as the training set and the remaining 20\% as the test set.
Different visible-channel combinations (four cases) were selected for MI/MM \cite{schalkBCI2000GeneralpurposeBraincomputer2004} and SEED \cite{zhengInvestigatingCriticalFrequency2015} using different SR scale factors (two, four, and eight).

\begin{table*}[htbp]
  \renewcommand{\arraystretch}{1.2}
  \centering
  \scriptsize
  \caption{Comparison with different hyperparameter configurations of ESTformer. AVG{\footnotesize $\pm$ STD} indicates average{\footnotesize $\pm$ standard} results of metrics.}
  \resizebox{\linewidth}{!}{
    \begin{tabular}{cccccccccccccc}
      \hline
      \hline
      \multirow{3}[0]{*}{No.} & \multicolumn{5}{c}{Param} & \multicolumn{8}{c}{Metrics (AVG {\footnotesize $\pm$ STD})}                                                                                                                                                                                                                                                                                                                                                               \\
      \cline{2-14}
                              & \multicolumn{3}{c}{Width} & \multicolumn{2}{c}{Depth}                           & \multicolumn{4}{c}{MI/MM} & \multicolumn{4}{c}{SEED}                                                                                                                                                                                                                                                                                                           \\
      \cline{2-14}
                              & {$\alpha_{\rm{t}}$}       & {$\alpha_{\rm{s}}$}                                 & {$r_{\rm{mlp}}$}       & {$L_{\rm{S}}$}           & {$L_{\rm{T}}$}        & NMSE $\downarrow$               & SNR $\uparrow$                  & PCC $\uparrow$                  & Minutes $\downarrow$             & NMSE $\downarrow$               & SNR $\uparrow$                  & PCC $\uparrow$                  & Minutes $\downarrow$
      \\
      \hline
      {0}                     & \textbf{0.60}             & \textbf{0.75}                                       & \textbf{4}             & \textbf{1}               & \textbf{1}            & \textbf{0.188{\footnotesize$\pm$0.001}} & \textbf{7.254{\footnotesize$\pm$0.017}} & \textbf{0.900{\footnotesize$\pm$0.000}} & \textbf{29.784{\footnotesize$\pm$0.240}} & \textbf{0.348{\footnotesize$\pm$0.001}} & \textbf{4.590{\footnotesize$\pm$0.014}} & \textbf{0.807{\footnotesize$\pm$0.001}} & \textbf{10.371{\footnotesize$\pm$0.189}}
      \\
      \hline
      1                       & \underline{0.30}          & \textcolor{mygray}{0.75}                            & \textcolor{mygray}{4}  & \textcolor{mygray}{1}    & \textcolor{mygray}{1} & 0.189{\footnotesize$\pm$0.001}          & 7.241{\footnotesize$\pm$0.017}          & 0.900{\footnotesize$\pm$0.000}          & 23.999{\footnotesize$\pm$1.012}          & 0.350{\footnotesize$\pm$0.001}          & 4.557{\footnotesize$\pm$0.014}          & 0.805{\footnotesize$\pm$0.001}          & 10.218{\footnotesize$\pm$0.148}
      \\
      2                       & \underline{0.90}          & \textcolor{mygray}{0.75}                            & \textcolor{mygray}{4}  & \textcolor{mygray}{1}    & \textcolor{mygray}{1} & 0.188{\footnotesize$\pm$0.001}          & 7.263{\footnotesize$\pm$0.020}          & 0.900{\footnotesize$\pm$0.000}          & 43.890{\footnotesize$\pm$0.254}          & 0.346{\footnotesize$\pm$0.001}          & 4.608{\footnotesize$\pm$0.014}          & 0.808{\footnotesize$\pm$0.001}          & 11.002{\footnotesize$\pm$0.166}
      \\
      \hline
      3                       & \textcolor{mygray}{0.60}  & \underline{0.50}                                    & \textcolor{mygray}{4}  & \textcolor{mygray}{1}    & \textcolor{mygray}{1} & 0.189{\footnotesize$\pm$0.001}          & 7.245{\footnotesize$\pm$0.019}          & 0.900{\footnotesize$\pm$0.000}          & 28.830{\footnotesize$\pm$0.276}          & 0.348{\footnotesize$\pm$0.001}          & 4.581{\footnotesize$\pm$0.014}          & 0.806{\footnotesize$\pm$0.001}          & 9.828{\footnotesize$\pm$0.214}
      \\
      4                       & \textcolor{mygray}{0.60}  & \underline{1.00}                                    & \textcolor{mygray}{4}  & \textcolor{mygray}{1}    & \textcolor{mygray}{1} & 0.189{\footnotesize$\pm$0.001}          & 7.225{\footnotesize$\pm$0.018}          & 0.899{\footnotesize$\pm$0.000}          & 30.458{\footnotesize$\pm$0.243}          & 0.347{\footnotesize$\pm$0.001}          & 4.595{\footnotesize$\pm$0.014}          & 0.807{\footnotesize$\pm$0.001}          & 10.409{\footnotesize$\pm$0.277}
      \\
      \hline
      5                       & \textcolor{mygray}{0.60}  & \textcolor{mygray}{0.75}                            & \underline{2}          & \textcolor{mygray}{1}    & \textcolor{mygray}{1} & 0.188{\footnotesize$\pm$0.001}          & 7.262{\footnotesize$\pm$0.016}          & 0.900{\footnotesize$\pm$0.000}          & 28.036{\footnotesize$\pm$0.381}          & 0.349{\footnotesize$\pm$0.001}          & 4.572{\footnotesize$\pm$0.014}          & 0.806{\footnotesize$\pm$0.001}          & 10.454{\footnotesize$\pm$0.334}
      \\
      6                       & \textcolor{mygray}{0.60}  & \textcolor{mygray}{0.75}                            & \underline{8}          & \textcolor{mygray}{1}    & \textcolor{mygray}{1} & 0.187{\footnotesize$\pm$0.001}          & 7.283{\footnotesize$\pm$0.018}          & 0.901{\footnotesize$\pm$0.000}          & 33.682{\footnotesize$\pm$0.241}          & 0.347{\footnotesize$\pm$0.001}          & 4.599{\footnotesize$\pm$0.013}          & 0.807{\footnotesize$\pm$0.001}          & 10.660{\footnotesize$\pm$0.284}
      \\
      \hline
      7                       & \textcolor{mygray}{0.60}  & \textcolor{mygray}{0.75}                            & \textcolor{mygray}{4}  & \underline{2}            & \textcolor{mygray}{1} & 0.187{\footnotesize$\pm$0.001}          & 7.289{\footnotesize$\pm$0.017}          & 0.901{\footnotesize$\pm$0.000}          & 44.245{\footnotesize$\pm$0.240}          & 0.347{\footnotesize$\pm$0.001}          & 4.598{\footnotesize$\pm$0.014}          & 0.807{\footnotesize$\pm$0.001}          & 14.184{\footnotesize$\pm$0.453}
      \\
      8                       & \textcolor{mygray}{0.60}  & \textcolor{mygray}{0.75}                            & \textcolor{mygray}{4}  & \underline{3}            & \textcolor{mygray}{1} & 0.186{\footnotesize$\pm$0.001}          & 7.295{\footnotesize$\pm$0.017}          & 0.901{\footnotesize$\pm$0.000}          & 58.588{\footnotesize$\pm$0.246}          & 0.347{\footnotesize$\pm$0.001}          & 4.596{\footnotesize$\pm$0.015}          & 0.807{\footnotesize$\pm$0.001}          & 18.193{\footnotesize$\pm$0.570}
      \\
      \hline
      9                       & \textcolor{mygray}{0.60}  & \textcolor{mygray}{0.75}                            & \textcolor{mygray}{4}  & \textcolor{mygray}{1}    & \underline{2}         & 0.188{\footnotesize$\pm$0.001}          & 7.266{\footnotesize$\pm$0.015}          & 0.900{\footnotesize$\pm$0.000}          & 39.631{\footnotesize$\pm$0.395}          & 0.347{\footnotesize$\pm$0.001}          & 4.599{\footnotesize$\pm$0.013}          & 0.807{\footnotesize$\pm$0.001}          & 12.131{\footnotesize$\pm$0.306}
      \\
      10                      & \textcolor{mygray}{0.60}  & \textcolor{mygray}{0.75}                            & \textcolor{mygray}{4}  & \textcolor{mygray}{1}    & \underline{3}         & 0.187{\footnotesize$\pm$0.001}          & 7.281{\footnotesize$\pm$0.016}          & 0.901{\footnotesize$\pm$0.000}          & 49.238{\footnotesize$\pm$0.313}          & 0.347{\footnotesize$\pm$0.001}          & 4.594{\footnotesize$\pm$0.015}          & 0.807{\footnotesize$\pm$0.001}          & 13.997{\footnotesize$\pm$0.486}
      \\
      \hline
      \hline
    \end{tabular}%
  }
  \label{tab:param}%
\end{table*}%

\subsection{Experiment Design}

First, we compared the proposed ESTformer with different hyperparameters in a preliminary study of the EEG SR task.
Subsequently, the main modules in ESTformer were compared to verify their effectiveness and efficiency in the EEG SR task.
Finally, experiments on two downstream tasks were conducted to explore the performance of the SR EEG data.

The details of these experiments are as follows.

\textbf{Experiment 1 - Hyperparameter comparison}.
The experiments were conducted under an SR scale factor of four and an LR channel combination setting for Case 1.
The experiments were repeated ten times using the MI/MM and SEED datasets.
We compared the main hyperparameter settings of ESTformer from two perspectives: model depth and model width.
Comparing each hyperparameter with the three settings, we trade-off efficiency and effectiveness to determine an appropriate combination of hyperparameters.
Additionally, the computational efficiency of ESTformer was compared with two EEG SR methods based on 2D-CNN to provide a broader assessment of its performance.

\textbf{Experiment 2 - SR Scaling Capability}.
The experiments were conducted across four SR scale factors (2, 4, 6, and 8), each paired with a specific LR channel combination setting for Case 1. 
To ensure robustness, each experiment was repeated ten times on the MI/MM and SEED datasets. 
In a preliminary study, the ESTformer model was trained under these conditions to assess its reconstruction capability at various SR scales.
We analyse the results in this part to examine the relationship between performance and SR scale factor adjustments.

\textbf{Experiment 3 - Ablation study}.
The experiments were conducted under all three SR scale factors and four LR channel combination settings.
The experiments were repeated ten times using the MI/MM and SEED datasets.
We compared the modified MAEs \cite{heMaskedAutoencodersAre2022}, proposed SIM, and proposed ESTformer to analyse the effectiveness of each module.
In addition, we adopted an easy-to-implement DeepCNN \cite{hanFeasibilityStudyEEG2018} as the enhancement module for the interpolated EEG data to validate whether ESTformer can replace traditional mathematical interpolation methods in a two-stage EEG SR manner.

\textbf{Experiment 4 - Downstream tasks}.
The experiments were conducted under all three SR scale factors and four LR channel combination settings.
The experiments were repeated five times using MI/MM and SEED datasets.
A simple 2-block MLP was chosen as the classifier, with the first block learning the feature dimension information and the second block learning the channel dimension representation.
Power spectral density (PSD) and differential entropy (DE) features were extracted for person identification and emotion-recognition tasks using MI/MM and SEED, respectively.
These features were extracted from the EEG LR, EEG GT, and EEG SR data interpolated by ESTformer.
We compared five individual bands and all the frequency bands of the extracted features to analyse the performance of the SR data in the frequency domain for downstream tasks.

We used the normalised mean square error (NMSE), signal-to-noise ratio (SNR), and Pearson correlation coefficient (PCC) as metrics for the EEG SR task.
The runtime in minutes is another important consideration in Ex. 1 and Ex. 2.
We used the classification accuracy (Acc) as a metric for the downstream tasks of EEG-based person identification and emotion recognition.
All these metrics can be calculated by averaging the results of the repeated experiments for all cases.

\begin{table}[htbp]
  \renewcommand{\arraystretch}{1.2}
  \centering
  \footnotesize
  \caption{Computation Efficiency comparisons between ESTformer and EEG SR models based on 2D-CNN.}
  \begin{tabular}{cccc}
    \hline
    \hline
    Methods                                             &   FLOPs (G)   & Params (M)    & Time (Hours) \\
    \hline
    ESTformer                                           & \textbf{1.5}  & 33.7          & \textbf{0.5} \\
    Generator \cite{corleyDeepEEGSuperresolution2018}   & 3332.2        & 16.5          & 9.0 \\
    Deep-CNN \cite{hanFeasibilityStudyEEG2018}          & 939.5         & \textbf{2.2}  & 2.3 \\
    \hline
    \hline
  \end{tabular}%
  \label{tab:Com-Eff}%
\end{table}%

\begin{figure*}[!htbp]
  \centering
  \includegraphics[width=1\textwidth]{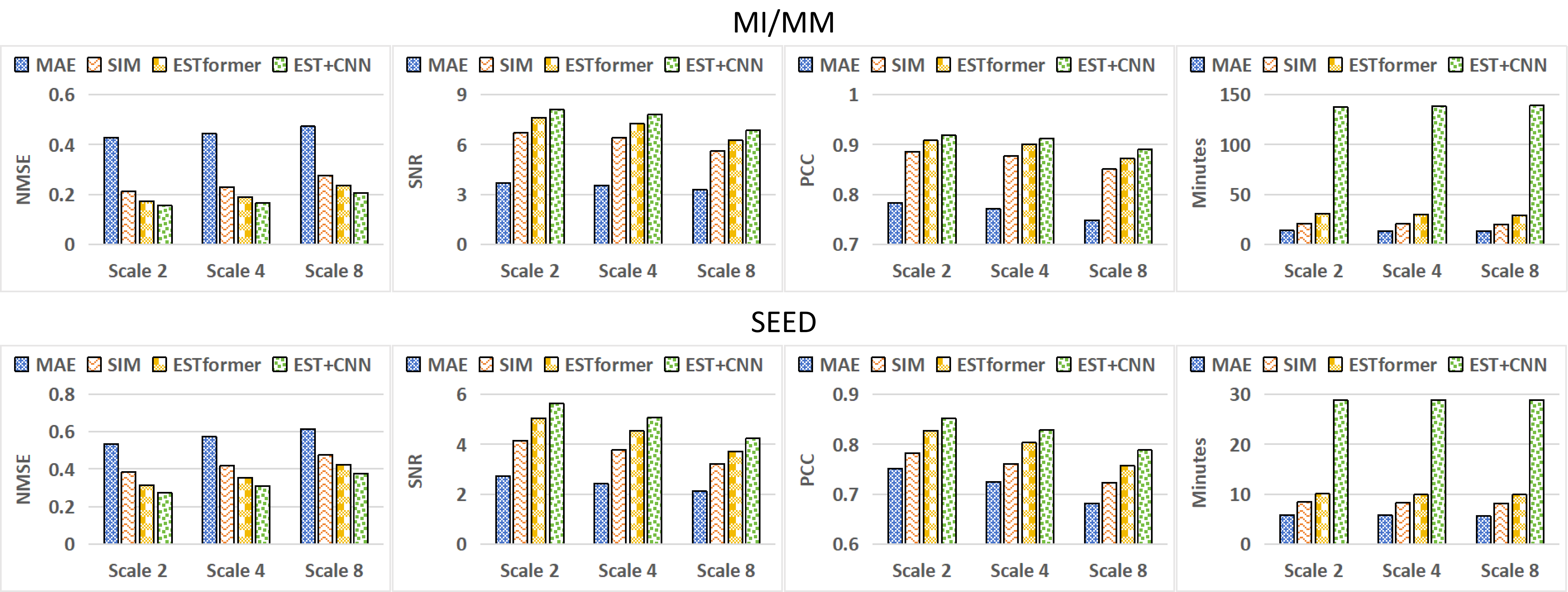}
  \caption{Comparisons between ESTformer and counterparts for ablation study.}
  \label{fig:ablation}
\end{figure*}

\begin{figure*}[!htpb]
  \centering
  \includegraphics[width=1\textwidth]{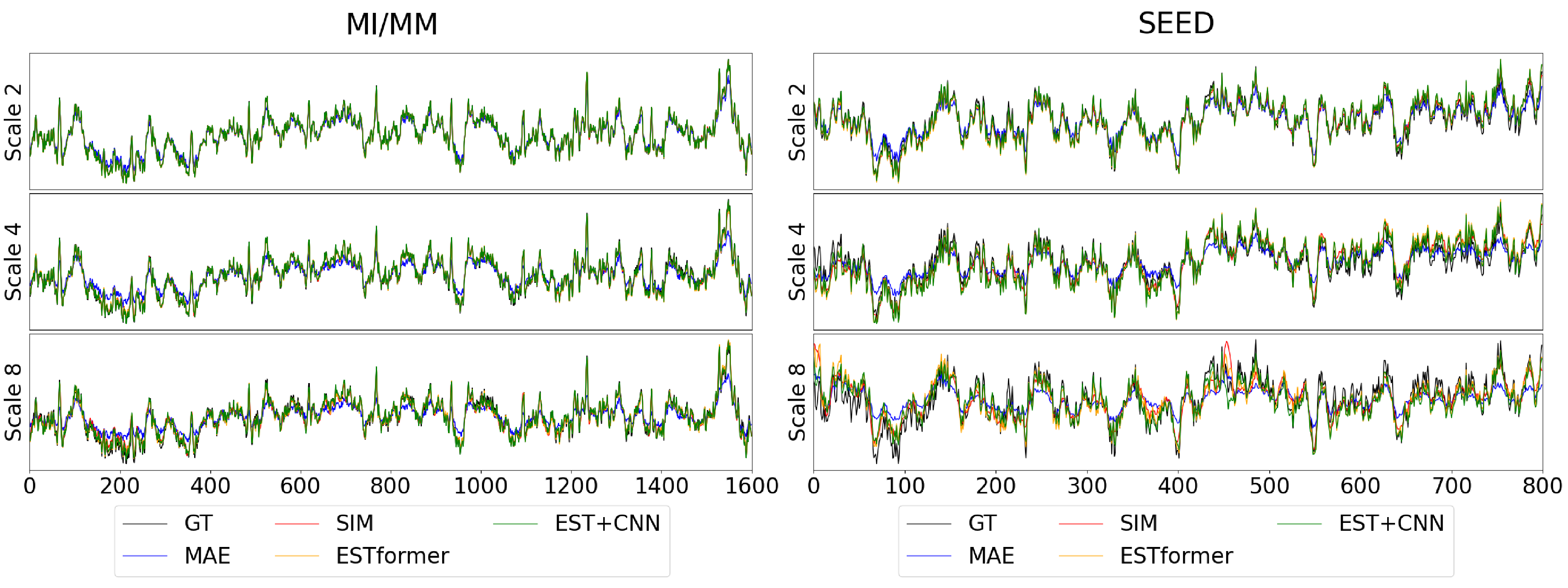}
  \caption{Time-axis visualisation results of EEG reconstruction in two datasets.}
  \label{fig:vist}
\end{figure*}

\begin{figure*}[!htpb]
  \centering
  \includegraphics[width=1\textwidth]{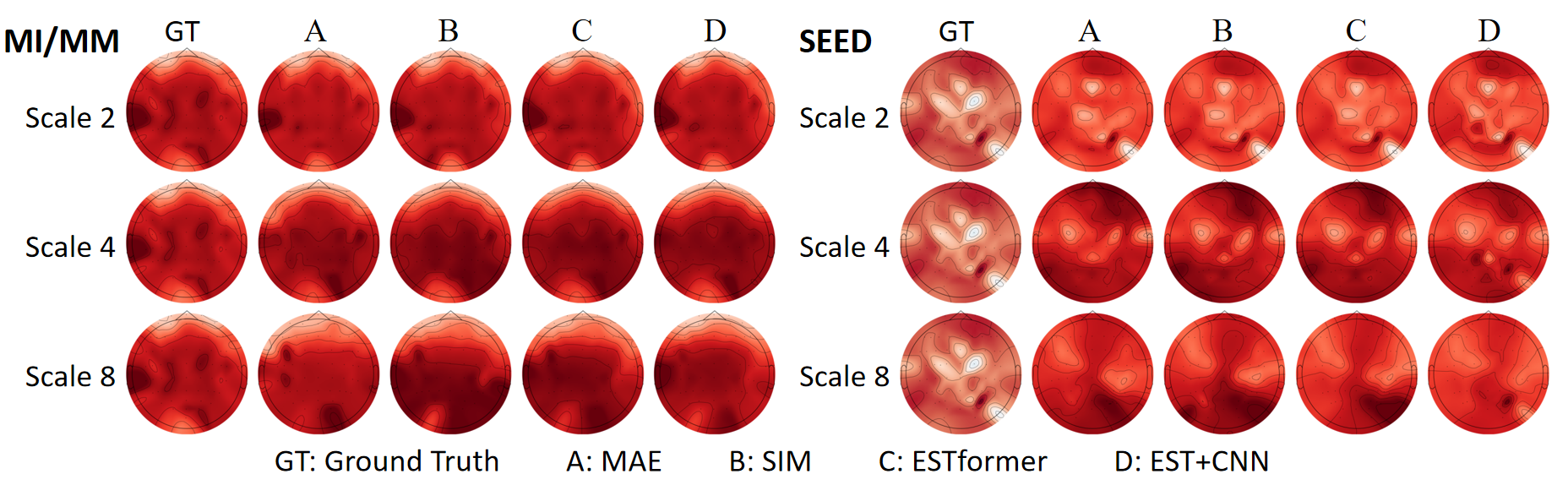}
  \caption{Channel-axis visualisation results of EEG reconstruction in two datasets.}
  \label{fig:visf}
\end{figure*}

\subsection{Hyperparameter Comparison\label{sec:ex_hyper}}

The setting of the main hyperparameters of the proposed model was based on two aspects: model depth and model width.
In the model width settings, $\alpha_{\rm{t}}$ represents the proportion of embedded feature length in the SIM module, $\alpha_{\rm{s}}$ represents the proportion of embedded feature length in the TRM module, and $r_{\rm{mlp}}$ represents the amplification ratio of the hidden layer in the MLP.
Specifically in model usage, $d_{\rm{s}}=C\times\alpha_{\rm{s}}$, $d_{\rm{t}}=T\times\alpha_{\rm{t}}$, where $d_{\rm{s}}$ denotes the length of the embedding from the spatial dimension in the TRM module, $d_{\rm{t}}$ denotes the length of the embedding from temporal dimension in the SIM module, and $C$ and $T$ respectively denote the number of electrode channels and the number of time-sampling points.
In the model depth settings, $L_{\rm{S}}$ represents the number of stacked layers of the CAB in SIM, and $L_{\rm{T}}$ represents the number of stacked layers of the TSAB in TRM.
Specifically in model usage, SIM uses a total of $2\times L_{\rm{S}}$ CAB, whereas TRM uses a total of $2\times L_{\rm{T}}$ TSAB.

The results of the hyperparameters comparison experiments are presented in \textbf{Table \ref{tab:param}}.
Bold font represents the default hyperparameter values as appropriate hyperparameter combinations: $\alpha_{\rm{t}}=0.6$, $\alpha_{\rm{s}}=0.75$, $L_{\rm{S}}=1$, $L_{\rm{T}}=1$, $r_{\rm{mlp}}=4$.
The underlined font represents the counterparts of specific hyperparameter.
When one parameter was adjusted, the other parameters used their default values.
It can be observed that although deepening the model depth can significantly improve the model performance, it also noticeably increases the computational time.
Therefore, the selection of hyperparameter values is based on the SR performance and computational time, considering the effects on both datasets.
AdamW was used as the optimiser.
The other main hyperparameter settings were a batch size of 36, dropout rate of 0.5, weight decay rate of 0.5, $\beta_1=0.9$, $\beta_2=0.95$, and learning rate of $5\times{10}^{-5}$.

As outlined in Sec.\ref{sec:3-1}, the use of self-attention mechanisms in the SIM and TRM modules is anticipated to improve computational efficiency over 2D-CNN-based architectures. 
To evaluate this, we compared ESTformer with two 2D-CNN-based methods: the DEEP-CNN approach from \cite{hanFeasibilityStudyEEG2018} and the DenseNet-like \cite{huangDenselyConnectedConvolutional2017} generator used in EEGSR-GAN \cite{corleyDeepEEGSuperresolution2018}.
We assessed the computational efficiency using three metrics: FLOPs per sample, total model parameters, and training time. 
As shown in \textbf{Table \ref{tab:Com-Eff}}, ESTformer achieved the lowest FLOPs and the shortest training time, aligned with our calculations in Sec.\ref{sec:3-1}, despite having a higher parameter count. 
This demonstrates that the self-attention mechanisms in ESTformer significantly contribute to its computational advantages over traditional 2D-CNN structures.

\begin{figure}[!htpb]
  \centering
  \includegraphics[width=1\columnwidth]{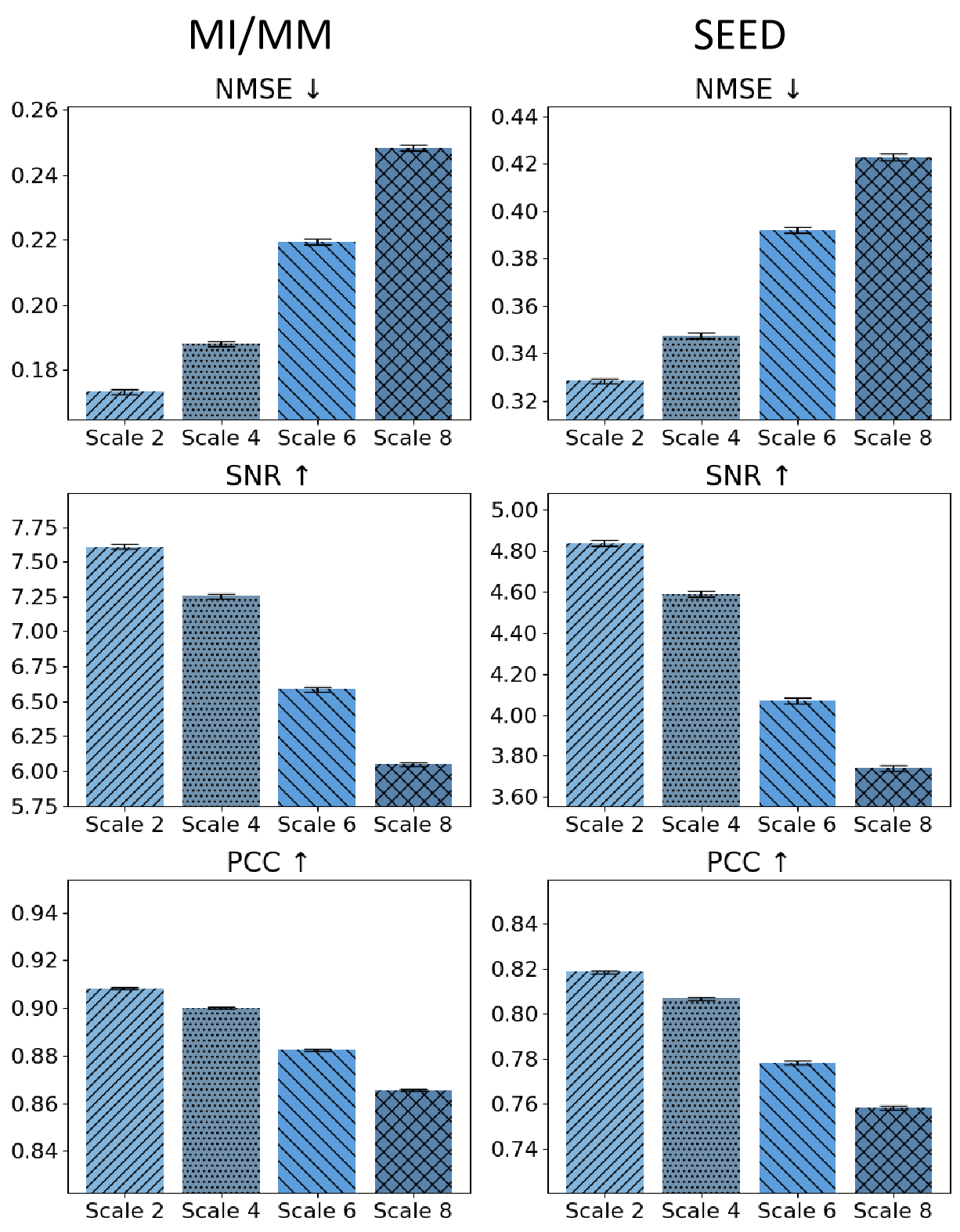}
  \caption{ESTformer performance at four SR scale settings in two datasets.}
  \label{fig:ex_mask}
\end{figure}

\subsection{SR scaling capability\label{sec:ex_scale}}
To assess the scaling capability of ESTformer preliminarily, we evaluated its performance across four scale factors (2, 4, 6, and 8) using two datasets.

As shown in \textbf{Fig.\ref{fig:ex_mask}}, ESTformer effectively reconstructs EEG data across varying scale factors. 
As anticipated, the performances of the three key metrics—NMSE, PCC, and SNR—decreases as the scale factor increases in both the MI/MM and SEED datasets. 
In addition, performance drops become more pronounced at higher scale factors, reflecting the challenge of accurately reconstructing EEG data when a larger proportion of channels are missing. 
This trend highlights the increased difficulty in maintaining reconstruction accuracy with fewer available data points at higher scaling levels.

Based on these observations, we focused on subsequent experiments with scale factors of 2, 4, and 8, aligned with the experimental settings in \cite{tangDeepEEGSuperresolution2023}.

\begin{table*}[htbp]
  \renewcommand{\arraystretch}{1.2}
  \centering
  \scriptsize
  \caption{Performance comparisons between ESTformer and state-of-the-art methods on MI/MM.
  ESTformer is compared with SI and EEGSR-GAN from an end-to-end aspect, while ESTformer+Deep-CNN is compared with AN+Deep-CNN and SI+Deep-EEGSR from a stage-to-stage aspect.}
      \centering
        \renewcommand{\arraystretch}{1.2}
\resizebox{\linewidth}{!}{
  \begin{tabular}{cccccccccc}
    \hline
    \hline
    \multirow{3}[0]{*}{Methods}                         & \multicolumn{9}{c}{Metrics(AVG{\footnotesize $\pm$ STD})}                                                                                                                                                                                                                                                                                       \\
    \cline{2-10}
                                                        & \multicolumn{3}{c}{scale factor 2}                & \multicolumn{3}{c}{scale factor 4} & \multicolumn{3}{c}{scale factor 8}                                                                                                                                                                                                             \\
    \cline{2-10}
                                                        & NMSE $\downarrow$                                 & SNR $\uparrow$                     & PCC $\uparrow$                     & NMSE $\downarrow$               & SNR $\uparrow$                  & PCC $\uparrow$                  & NMSE $\downarrow$               & SNR $\uparrow$                  & PCC $\uparrow$                  \\
    \hline
    SI \cite{freedenSphericalSplineInterpolation1984}   & 0.287{\footnotesize$\pm$0.000}                            & 5.417{\footnotesize$\pm$0.000}             & 0.855{\footnotesize$\pm$0.000}             & 0.317{\footnotesize$\pm$0.000}          & 4.993{\footnotesize$\pm$0.000}          & 0.839{\footnotesize$\pm$0.000}          & 0.367{\footnotesize$\pm$0.000}          & 4.351{\footnotesize$\pm$0.000}          & 0.811{\footnotesize$\pm$0.000}
    \\
    EEGSR-GAN \cite{corleyDeepEEGSuperresolution2018}   & 0.224{\footnotesize$\pm$0.000}                            & 6.496{\footnotesize$\pm$0.008}             & 0.887{\footnotesize$\pm$0.001}             & 0.258{\footnotesize$\pm$0.002}          & 5.883{\footnotesize$\pm$0.039}          & 0.860{\footnotesize$\pm$0.001}          & 0.308{\footnotesize$\pm$0.003}          & 5.110{\footnotesize$\pm$0.049}          & 0.832{\footnotesize$\pm$0.003}
    \\
    \textbf{ESTformer}                                  & \textbf{0.174{\footnotesize$\pm$0.004}}                   & \textbf{7.598{\footnotesize$\pm$0.099}}    & \textbf{0.908{\footnotesize$\pm$0.002}}    & \textbf{0.189{\footnotesize$\pm$0.002}} & \textbf{7.239{\footnotesize$\pm$0.045}} & \textbf{0.900{\footnotesize$\pm$0.001}} & \textbf{0.237{\footnotesize$\pm$0.007}} & \textbf{6.252{\footnotesize$\pm$0.132}} & \textbf{0.872{\footnotesize$\pm$0.004}}
    \\
    \hline
    AN+Deep-CNN \cite{hanFeasibilityStudyEEG2018}       & 0.206{\footnotesize$\pm$0.019}                            & 6.864{\footnotesize$\pm$0.406}             & 0.900{\footnotesize$\pm$0.007}             & 0.231{\footnotesize$\pm$0.021}          & 6.378{\footnotesize$\pm$0.385}          & 0.879{\footnotesize$\pm$0.012}          & 0.276{\footnotesize$\pm$0.044}          & 5.620{\footnotesize$\pm$0.663}          & 0.855{\footnotesize$\pm$0.019}
    \\
    SI+Deep-EEGSR \cite{tangDeepEEGSuperresolution2023} & 0.165{\footnotesize$\pm$0.001}                            & 7.816{\footnotesize$\pm$0.016}             & \textbf{0.920{\footnotesize$\pm$0.000}}    & 0.177{\footnotesize$\pm$0.001}          & 7.514{\footnotesize$\pm$0.021}          & 0.907{\footnotesize$\pm$0.000}          & 0.214{\footnotesize$\pm$0.002}          & 6.688{\footnotesize$\pm$0.034}          & 0.885{\footnotesize$\pm$0.001}
    \\
    \textbf{ESTformer+Deep-CNN}                         & \textbf{0.155{\footnotesize$\pm$0.003}}                   & \textbf{8.102{\footnotesize$\pm$0.092}}    & 0.918{\footnotesize$\pm$0.002}             & \textbf{0.166{\footnotesize$\pm$0.002}} & \textbf{7.790{\footnotesize$\pm$0.055}} & \textbf{0.912{\footnotesize$\pm$0.001}} & \textbf{0.207{\footnotesize$\pm$0.006}} & \textbf{6.852{\footnotesize$\pm$0.133}} & \textbf{0.890{\footnotesize$\pm$0.004}}
    \\
    \hline
    \hline
      \end{tabular}
      }
        \label{tab:SOTA-MI}%
\end{table*}%

\begin{table*}[htbp]
  \renewcommand{\arraystretch}{1.2}
  \centering
  \scriptsize
  \caption{Performance comparisons between ESTformer and state-of-the-art methods on SEED.
  ESTformer is compared with SI and EEGSR-GAN from an end-to-end aspect, while ESTformer+Deep-CNN is compared with AN+Deep-CNN and SI+Deep-EEGSR from a stage-to-stage aspect.}
      \centering
        \renewcommand{\arraystretch}{1.2}
\resizebox{\linewidth}{!}{
  \begin{tabular}{cccccccccc}
    \hline
    \hline
    \multirow{3}[0]{*}{Methods}                         & \multicolumn{9}{c}{Metrics(AVG{\footnotesize $\pm$ STD})}                                                                                                                                                                                                                                                                                       \\
    \cline{2-10}
                                                        & \multicolumn{3}{c}{scale factor 2}                & \multicolumn{3}{c}{scale factor 4} & \multicolumn{3}{c}{scale factor 8}                                                                                                                                                                                                             \\
    \cline{2-10}
                                                        & NMSE $\downarrow$                                 & SNR $\uparrow$                     & PCC $\uparrow$                     & NMSE $\downarrow$               & SNR $\uparrow$                  & PCC $\uparrow$                  & NMSE $\downarrow$               & SNR $\uparrow$                  & PCC $\uparrow$                  \\
    \hline
    SI \cite{freedenSphericalSplineInterpolation1984}   & 0.684{\footnotesize$\pm$0.000}                            & 1.647{\footnotesize$\pm$0.000}             & 0.585{\footnotesize$\pm$0.000}             & 0.703{\footnotesize$\pm$0.000}          & 1.532{\footnotesize$\pm$0.000}          & 0.567{\footnotesize$\pm$0.000}          & 0.755{\footnotesize$\pm$0.000}          & 1.219{\footnotesize$\pm$0.000}          & 0.527{\footnotesize$\pm$0.000}          \\
    EEGSR-GAN \cite{corleyDeepEEGSuperresolution2018}   & 0.514{\footnotesize$\pm$0.002}                            & 2.892{\footnotesize$\pm$0.019}             & 0.693{\footnotesize$\pm$0.001}             & 0.597{\footnotesize$\pm$0.002}          & 2.241{\footnotesize$\pm$0.013}          & 0.629{\footnotesize$\pm$0.001}          & 0.652{\footnotesize$\pm$0.001}          & 1.855{\footnotesize$\pm$0.006}          & 0.583{\footnotesize$\pm$0.001}          \\
    \textbf{ESTformer}                                  & \textbf{0.315{\footnotesize$\pm$0.014}}                   & \textbf{5.027{\footnotesize$\pm$0.199}}    & \textbf{0.827{\footnotesize$\pm$0.009}}    & \textbf{0.353{\footnotesize$\pm$0.007}} & \textbf{4.528{\footnotesize$\pm$0.084}} & \textbf{0.803{\footnotesize$\pm$0.004}} & \textbf{0.425{\footnotesize$\pm$0.011}} & \textbf{3.713{\footnotesize$\pm$0.113}} & \textbf{0.757{\footnotesize$\pm$0.007}} \\
    \hline
    AN+Deep-CNN \cite{hanFeasibilityStudyEEG2018}       & 0.485{\footnotesize$\pm$0.005}                            & 3.143{\footnotesize$\pm$0.047}             & 0.713{\footnotesize$\pm$0.004}             & 0.608{\footnotesize$\pm$0.006}          & 2.159{\footnotesize$\pm$0.045}          & 0.622{\footnotesize$\pm$0.006}          & 0.672{\footnotesize$\pm$0.007}          & 1.724{\footnotesize$\pm$0.044}          & 0.569{\footnotesize$\pm$0.002}          \\
    SI+Deep-EEGSR \cite{tangDeepEEGSuperresolution2023} & 0.413{\footnotesize$\pm$0.010}                            & 3.843{\footnotesize$\pm$0.108}             & 0.762{\footnotesize$\pm$0.007}             & 0.524{\footnotesize$\pm$0.002}          & 2.806{\footnotesize$\pm$0.018}          & 0.682{\footnotesize$\pm$0.002}          & 0.580{\footnotesize$\pm$0.005}          & 2.366{\footnotesize$\pm$0.038}          & 0.641{\footnotesize$\pm$0.003}          \\
    \textbf{ESTformer+Deep-CNN}                         & \textbf{0.274{\footnotesize$\pm$0.014}}                   & \textbf{5.634{\footnotesize$\pm$0.223}}    & \textbf{0.852{\footnotesize$\pm$0.008}}    & \textbf{0.311{\footnotesize$\pm$0.005}} & \textbf{5.072{\footnotesize$\pm$0.076}} & \textbf{0.829{\footnotesize$\pm$0.003}} & \textbf{0.377{\footnotesize$\pm$0.008}} & \textbf{4.233{\footnotesize$\pm$0.094}} & \textbf{0.788{\footnotesize$\pm$0.005}} \\
    \hline
    \hline
  \end{tabular}%
      }
        \label{tab:SOTA-SEED}%
\end{table*}%

\subsection{Ablation Study\label{sec:ex_ablation}}

Based on the set of hyperparameters obtained using Eqs. 1, The following models were selected for the ablation study: MAEs \cite{heMaskedAutoencodersAre2022}, SIM, ESTformer ( SIM and TRM), and ESTformer+Deep-CNN (denoted as EST+CNN in the figures), which replaced the mathematical interpolation method of averaging neighbour channels (AN) with the proposed ESTformer.
We modified MAEs \cite{heMaskedAutoencodersAre2022} based on the characteristics of the EEG data and the requirements of the EEG SR task.
Specifically, the modified MAEs are composed of an SSAB with 1D positional encoding using a fixed-mask sampling strategy.

The comparative results are shown in \textbf{Fig.\ref{fig:ablation}}.
Compared with modified MAEs, the proposed SIM uses 3D positional encoding and CAB, resulting in a significant improvement.
ESTformer, built on SIM and TRM, outperforms SIM, indicating that additional temporal reconstruction is needed for modelling long-term time dependency through SIM.
ESTformer+Deep-CNN (EST+CNN) applies a Deep-CNN \cite{hanFeasibilityStudyEEG2018} to the interpolated EEG data obtained from ESTformer, resulting in improved but significant time consumption, as mentioned in \ref{sec:3-1}.

Accordingly, \textbf{Fig.\ref{fig:vist}} and \textbf{Fig.\ref{fig:visf}} show the reconstructed EEG time series and scalp potentials, respectively.
In \textbf{Fig.\ref{fig:vist}}, EEG time series (averaged between channels) by ESTformer (in orange) better fits the GT (in black) than the SIM (in red) in time-axis, which demonstrates TRM is effective in modelling the long-term time dependency.
In \textbf{Fig.\ref{fig:visf}}, we calculate PSD features in MI/MM and DE features in SEED, aligning to downstream task settings in Sec.\ref{sec:ex_downstream}. 
The distributions of scalp potentials (averaged between frequency bands) obtained by ESTformer and ESTformer+Deep-CNN (EST+CNN) were also closer to the GT than their counterparts, especially for the outline of the critical location.

\subsection{Overall Comparison with Existing Methods\label{sec:ex_overall}}

Following the ablation study, ESTformer is considered a one-stage approach and ESTformer+Deep-CNN is considered a two-stage approach.
From this perspective, we compared existing state-of-the-art EEG SR methods.
ESTformer was compared with SI and EEGSR-GAN, whereas ESTformer+Deep-CNN was compared with AN+Deep-CNN and SI+Deep-EEGSR \cite{tangDeepEEGSuperresolution2023}.

\textbf{Table \ref{tab:SOTA-MI}} and \textbf{Table \ref{tab:SOTA-SEED}} present the results in the MI/MM and the SEED, respectively.
In these two datasets, the proposed ESTformer shows a significant improvement in one-stage approaches, whereas ESTformer+Deep-CNN can also outperform the state-of-the-art method in two-stage approaches.
AN+Deep-CNN \cite{hanFeasibilityStudyEEG2018} outperforms EEGSR-GAN in MI/MM, whereas EEGSR-GAN achieves competitive results with AN+Deep-CNN \cite{hanFeasibilityStudyEEG2018} in SEED.
This finding indicates that modelling challenging data in an end-to-end manner may be a better choice, conforming to the notable performance of ESTformer compared with SI+Deep-EEGSR \cite{tangDeepEEGSuperresolution2023} in the SEED.
Overall, these results indicate that the transformer-based ESTformer can effectively and efficiently model challenging data.

\begin{table}[htbp]
  \renewcommand{\arraystretch}{1.2}
  \centering
  \scriptsize
  \caption{Performance comparisons between SR data and others for EEG-based person identification tasks on the MI/MM.}
  \begin{tabular}{cccccccc}
    \hline
    \hline
    \multicolumn{2}{c}{\multirow{2}[0]{*}{Scale}} & \multicolumn{6}{c}{Acc $\uparrow$}                                                                \\
    \cline{3-8}
                              &                         & Delta                              & Theta  & Alpha  & Beta            & Gamma           & All    \\
    \hline
    {1}                       & {GT}                    & 77.791                             & 88.037 & 93.367 & 93.485          & 96.802          & 89.517 \\
    \hline
    \multirow{2}[0]{*}{2}     & LR                      & 67.190                             & 74.511 & 82.159 & 88.288          & 90.761          & 87.681 \\
                              & SR                      & 75.800                             & 82.203 & 90.228 & 93.730          & \textbf{94.611} & 90.139 \\
    \hline
    \multirow{2}[0]{*}{4}     & LR                      & 51.029                             & 57.810 & 67.634 & 75.229          & 78.013          & 71.980 \\
                              & SR                      & 72.416                             & 77.162 & 85.113 & 88.185          & \textbf{90.280} & 85.423 \\
    \hline
    \multirow{2}[0]{*}{8}     & LR                      & 27.924                             & 32.129 & 37.807 & 43.715          & 49.252          & 43.182 \\
                              & SR                      & 63.844                             & 66.338 & 74.859 & \textbf{81.863} & 81.056          & 78.124 \\
    \hline
    \hline
  \end{tabular}%
  \label{tab:CLS-MI}%
\end{table}%

\begin{table}[htbp]
  \renewcommand{\arraystretch}{1.2}
  \centering
  \scriptsize
  \caption{Performance comparisons between SR data and others for EEG-based emotion recognition tasks on the SEED.}
  \begin{tabular}{cccccccc}
    \hline
    \hline
    \multicolumn{2}{c}{\multirow{2}[0]{*}{Scale}} & \multicolumn{6}{c}{Acc $\uparrow$}                                                       \\
    \cline{3-8}
                              &                         & Delta                              & Theta  & Alpha  & Beta   & Gamma  & All             \\
    \hline
    {1}                       & {GT}                    & 50.369                             & 53.246 & 54.605 & 55.556 & 57.020 & 69.279          \\
    \hline
    \multirow{2}[0]{*}{2}     & LR                      & 46.490                             & 49.109 & 49.376 & 51.940 & 52.052 & 62.154          \\
                              & SR                      & 49.990                             & 53.276 & 55.110 & 55.364 & 57.278 & \textbf{66.066} \\
    \hline
    \multirow{2}[0]{*}{4}     & LR                      & 42.798                             & 44.745 & 45.012 & 48.466 & 48.143 & 54.965          \\
                              & SR                      & 47.615                             & 50.630 & 52.299 & 53.497 & 54.826 & \textbf{62.015} \\
    \hline
    \multirow{2}[0]{*}{8}     & LR                      & 40.980                             & 41.746 & 42.204 & 45.134 & 43.672 & 49.294          \\
                              & SR                      & 46.262                             & 48.027 & 50.036 & 51.999 & 51.814 & \textbf{56.539} \\
    \hline
    \hline
  \end{tabular}%
  \label{tab:CLS-SEED}%
\end{table}%

\subsection{Downstream Tasks\label{sec:ex_downstream}}

We conducted experiments on downstream tasks of person identification and emotion recognition.
We compared the ESTformer-derived full-channel SR, visible-channel LR, and original full-channel GT data.
According to common EEG classification tasks, PSD features are extracted for person identification and DE features are extracted for emotion recognition.
PSD and DE features were divided into five frequency bands (Delta, Theta, Alpha, Beta, and Gamma) with dimensions of $R^{C\times1}$.
The effect of using features from all five frequency bands with dimensions of $R^{C\times5}$ was also considered.

\textbf{Table \ref{tab:CLS-MI}} and \textbf{Table \ref{tab:CLS-SEED}} represent the results of the person identification task and the emotion recognition task, respectively.
Bold font represents the best performance for the same SR scale factor setting.
Overall, the SR data consistently outperformed the LR data and, in some cases, achieved competitive performance relative to the GT data, particularly when the scale factor was two. This suggests that the ESTformer-derived interpolated data may introduce features that are more "learnable" for downstream models, providing benefits over the original full-channel data in certain scenarios.

For the person identification task, the SR data demonstrated superior performance, particularly in high-frequency bands. This is likely because of the nature of motor MI tasks, which are known to elicit neural activity in the high-frequency range, making SR data more effective at capturing relevant discriminative features in this context. The improvements ranged significantly from 2\% to 38\%, highlighting the potential of SR data to enhance person identification performance, particularly with larger SR scale factors.

By contrast, the emotion-recognition task showed modest improvements with the SR data, with gains ranging from 3\% to 8\% across the different frequency bands. Interestingly, the best performance was achieved when using features from all five frequency bands, emphasising the complexity of affective computing tasks. The smaller improvements in this task underscore the challenge of capturing the nuanced neural signals associated with emotional states and suggest that while SR data can be beneficial, the task requires leveraging the full spectrum of EEG signals for optimal performance.

These findings have broad implications for EEG-based applications. The ability of ESTformer to supplement information from missing channels indicates that lightweight EEG devices with fewer channels can potentially be used for real-time applications, with SR methods compensating for missing data. This is particularly relevant for EEG-based biometric systems, where SR data show significant promise for enhancing person identification. In this context, a multichannel device can be used to collect high-quality data for training (as GT data), whereas a more practical, fewer-channel device can be employed in real-time settings (with LR data interpolated to SR data for testing), enabling more accessible and efficient EEG data collection without compromising performance.

In conclusion, ESTformer offers a powerful solution for improving the usability of low-channel EEG data across a range of downstream tasks, facilitating the development of lightweight EEG acquisition systems while maintaining robust performance.

\section{Conclusion and Future Work\label{sec:5}}

In this study, we introduced ESTformer, an end-to-end transformer-based framework designed to efficiently reconstruct high-resolution (HR) EEG data by leveraging the spatiotemporal dependencies inherent in EEG signals. The ESTformer architecture integrates a SIM and TRM that employ SSA and TSA, respectively, along with 3D spatial and 1D temporal positional encoding techniques.

Our experimental results indicate that the following: (1) ESTformer achieves state-of-the-art performance in EEG signal reconstruction; (2) the reconstructed (SR) EEG data significantly enhance performance in downstream EEG-based tasks, including person identification and emotion recognition.

However, some limitations remain: (1) As a transformer-based framework, ESTformer remains computationally intensive, which could limit its applicability in real-time scenarios; (2) the ESTformer architecture uses a straightforward cascade of SIM and TRM modules, which may benefit from alternative designs that facilitate more complex interactions between modules to improve learning efficiency.

In addition, several directions warrant further investigation.

(1) Performance under Complex Scenarios: Although ESTformer exhibits robust performance, its effectiveness under challenging conditions—such as cross-subject variations, task diversity, and long-term data acquisition—requires further generalisation validation. Thus, more adaptive and resilient modelling strategies should be explored.

(2) Comparative Analyses with Self-Supervised Representations and Privacy Considerations: The advantages of representations generated by SR relative to those learned through self-supervised approaches remain underexplored. Furthermore, ensuring privacy in the reconstructed EEG data is crucial, especially given the sensitivity of personal neurophysiological data. 

(3) Exploration of multimodal synergies: Given the unique strengths of different modalities across various resolutions, a multimodal approach holds the potential to enhance EEG reconstruction quality. By leveraging modality-specific advantages, future research can provide an enriched synergistic framework for EEG analysis with fidelity.

(4) Application of Implicit Neural Representations: The adoption of implicit neural representation techniques may further enable bidirectional mapping between EEG source activity and scalp electrode signals, facilitating flexible EEG signal reconstruction. The incorporation of these methods may provide a more versatile solution that accommodates diverse reconstruction demands.

\ifCLASSOPTIONcompsoc
  \section*{Acknowledgments}
\else
  \section*{Acknowledgment}
\fi
This study was supported by the National Natural Science Foundation of China under grant numbers 61806078, 62076094, and 61976091.

\section*{Declaration of Competing Interest}
The authors declare that they have no competing financial interests or personal relationships that may have influenced the work reported in this paper.


\bibliographystyle{IEEEtran}
\normalem
\bibliography{ref}

\end{document}